\documentclass[10pt, conference, compsocconf]{IEEEtran}

\ifCLASSINFOpdf

\else

\fi

\ifCLASSINFOpdf
\usepackage[pdftex]{graphicx}
\graphicspath{{./graph/}}
\DeclareGraphicsExtensions{.pdf,.jpeg,.png}
\else
\usepackage[dvips]{graphicx}
\graphicspath{{./graph/}}
\DeclareGraphicsExtensions{.eps}
\fi

\usepackage{algorithm}
\usepackage[vlined, ruled, algo2e]{algorithm2e}

\usepackage{amssymb}
\usepackage{amsmath}
\usepackage[table,xcdraw]{xcolor}
\usepackage{booktabs}

\begin{document}
%
\title{Extracting Point of Interest and Classifying Environment for Low Sampling Crowd Sensing Smartphone Sensor Data\vspace{-0.7cm}}


\author{\IEEEauthorblockN{Billy Pik Lik Lau\IEEEauthorrefmark{1}, 
		Marakkalage Sumudu Hasala\IEEEauthorrefmark{1},
		Viswanath Sanjana Kadaba\IEEEauthorrefmark{1}, \\
		Balasubramaniam Thirunavukarasu\IEEEauthorrefmark{2},
		Chau Yuen\IEEEauthorrefmark{1},
		Belinda Yuen\IEEEauthorrefmark{1},
		Richi Nayak\IEEEauthorrefmark{2}
		}
\IEEEauthorblockA{
	  Engineering Product Development, Singapore University of Technology and Design, Singapore \IEEEauthorrefmark{1}\\
	  School of Electrical Engineering and Computer Science, Queensland University of Technology, Australia
	  \IEEEauthorrefmark{2} \\
	  Email: billy\_lau@sutd.edu.sg, 
	  marakkalage@mymail.sutd.edu.sg,
	  yuenchau@sutd.edu.sg
}
\vspace{-0.5cm}}

\maketitle

\begin{abstract}
The advancement of smartphones with various type of sensors enabled us to harness diverse information with crowd sensing mobile application. 
However, traditional approaches have suffered drawbacks such as high battery consumption as a trade off to obtain high accuracy data using high sampling rate.
To mitigate the battery consumption, we proposed low sampling point of interest (POI) extraction framework, which is built upon validation based stay points detection (VSPD) and sensor fusion based environment classification (SFEC).
We studied various of clustering algorithm and showed that density based spatial clustering of application with noise (DBSCAN) algorithms produce most accurate result among existing methods.
The SFEC model is utilized for classifying the indoor or outdoor environment of the POI clustered earlier by VSPD.
Real world data are collected, benchmarked using existing clustering method to denote effectiveness of low sampling rate model in high noise spatial temporal data.
\end{abstract}

\begin{IEEEkeywords}
Big Data, POI, Clustering, Environment Classification, Trajectory, Sensor Fusion
\end{IEEEkeywords}

\vspace{-0.20cm}
\section{Introduction}
\label{sec_Introduction}

Data collected using mobile devices such as smart phone and wearable devices have been explicitly studied by various researchers in \cite{hoteit2014estimating,kang2013exploring} as they are able to generate various insights about a particular user life pattern. 
Among all the information captured, point of interest (POI), user mobility, user life pattern, and transportation mode gained a lot of attention; but major existing crowd sensing applications suffer from the drawbacks of high battery consumption. 
The reason behind high battery consumption is due to high sampling rate as well as limited battery capacity of mobile devices, which render most of the methods only suitable for short term data collection.

To solve the aforementioned challenges, we proposed \textit{ultra low} sampling architecture for extracting POI utilizing various sensor information available on a smart-phone. 
One of the main challenges of ultra low sampling architecture is trade off for accuracy in order to obtain data at a very long interval compared to traditional method. 
Another problem that may arise is highly correlated with the data quality, which may produce inaccurate output after computation.

The main objective of the paper is to study POI based on the various sensor information generated by smart phone using low sampling rate. 
Such technique will find application in better understanding user life pattern, which in return can improve user life style.
However, in a crowd sourcing / sensing environment, there are a wide variety of devices by different users, and it is not possible to retrieve ground truth from the users.
In this paper, our main focus is to verify the effectiveness of clustering algorithm based on data collected from volunteer users.
Our model require minimal attention and is suitable for helping elderly to understand their life pattern after retirements. These insights can be used for the authorities to improve certain location facilities, which will be reported in the continued works.

Various information has been collected through the application we have previously developed in \cite{viswanath2014smart}.
Data collected from smart-phone are uploaded to the server and off-line analysis are performed to gain insights about POI and trajectory. 
We first extract data from database, and later perform several preprocessing techniques namely \textit{denoise} and \textit{timeSync}. 
The former mentioned technique allows us to remove any potential noise, outliers, and duplicate data; whereas the latter mentioned technique is used for synchronizing the data with different timestamps. 
Subsequently, we apply VSPD algorithm to extract the POI, while the DBSCAN is used for clustering similar POI. 
Trajectory can be obtained through labeling each POI with the time stamp based on duration of each stay point. 
As our main concern is to determine POI, we further study the characteristic of the POI and classify it into indoor vs outdoor, or private vs public using SFEC technique. The SFEC is implemented using multi sensor data obtained from mobile devices.
\vspace{-0.05cm}

Our contributions in this paper are listed as follows:
\begin{itemize}
	\item We present VSPD algorithm, which is capable to distinguish valid stay points due to poor the data quality of low sampling rate. Stay points are validated using the proposed module and later clustered into POI based on temporal data.	
	\item SFEC method is presented to distinguish POI by its indoor and outdoor characteristic. In addition, we also categorize the POI by its property, whether it is private or public premises. 
\end{itemize}

The rest of the paper is structured as follows: In Section \ref{sec_SystemArchitecture}, we propose ultra low sampling architecture and discuss about the data processing techniques. In Section \ref{sec_POIExtraction}, VSPD algorithm is introduced and benchmarked against existing clustering and stay point detection algorithms. After clustering similar POI, we performed environment classification using SFEC method and show the efficiency of our methods in Section \ref{sec_SFEC}. Finally, we conclude our work in Section \ref{sec_conclusion}.

\vspace{-0.15cm}
\section{System Architecture and Framework}
\label{sec_SystemArchitecture}
In this section, we present the ultra low sampling rate architecture for data collection and processing.
\vspace{-0.15cm}

\subsection{User Trajectory and Point of Interest Detection}
\label{subsec_userTrajectory}
User mobility has been studied in \cite{li2008mining,lou2009map} and several features extraction show the necessity of extracting stay points before obtaining insights from user trajectory. Stay points are known as POI in our study scenario, where it can be a particular user home, work place, leisure spot, and etc. 

To study the POI, trajectory data and user mobility are often used to visualize the path a user takes from one POI to another.
Extracting trajectory features yields a lot of important information such as travel distance, duration, POI location, transportation mode, and etc.
On the other hand, transportation mode (TM) is identified through the feature extraction using methods such as probability estimation \cite{zheng2008learning}. 
Other related works such as \cite{hoteit2014estimating, zheng2010geolife} use mobile phone as platform for data collection and try to capture trajectory data using various interpolation methods (linear, cubic, and nearest). 
Finer detail of TM, such as motion patterns have been studied in \cite{suzuki2007learning}, which uses Hidden Markov Model (HMM) and $k$-means clustering for obtaining motion patterns; then it is also serve as anomaly detection. 

However, most of the methods only deal with high sampling rate of data and are not suitable for day long data collection due to limited battery capacity.
We incorporated validation function for the stay point detection in order to validate low sampling rate of data. 
\vspace{-0.15cm}

\subsection{System Architecture and Data Processing}
\label{subsec_systemArchitecture}
The system framework for collecting POI information can be divided into two phases, which are data collection and data processing. 
Data collection phase involves installation of mobile application at volunteers' smart phone. 
Data collected from mobile devices, which are (1) Location information, (2) Social Information, and (3) Activity Information. 
The location information consists of user geographical positioning data, while the social information is composed of sound information collected from the mobile devices microphone. 
The activity information consists of data such as TM and type of activity that users are engaged in.

\begin{figure}[htb]
	\centering
	\includegraphics[width=0.46\textwidth]{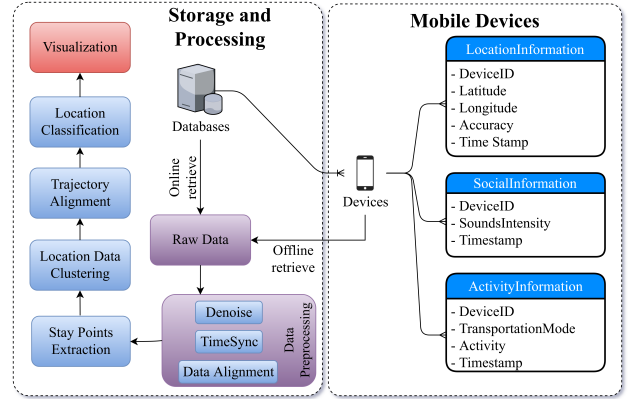}	
	\vspace{-1.0em}
	\centering \caption{Overview of the system architecture}
	\label{fig_systemArchitecture}
	\vspace{-1.5em}
\end{figure}

We have previously developed the application using Android frameworks in \cite{viswanath2014smart} and is used for data collection in this paper.
It has been further fine-tuned to adapt the sampling rate of 5 minutes and optimized for off-line data collection, when there is no real-time Internet connection.
The overview of the system architecture is shown in Fig. \ref{fig_systemArchitecture}. 

The preprocessing of data is performed in order to generate usable data to extract useful features. 
First, the raw data will be passed through \textit{denoise} process to remove any duplicate records. 
The denoise module checks the data time-stamp and eliminates data with same time-stamp. 
The reason for duplicated data is mostly due to the redundant data kept by mobile phone, which fails to receive acknowledgment from server upon upload.
The timeSync module act as the function for linking different data to a similar time stamp in order to perform data fusion computation, which will be presented in later part.
Data alignment starts by aligning the data from different database based on the synchronized time for each dataset into single matrices. 
The overall data processing is shown in the Algorithm \ref{algo_SystemOverallAlgo}.

\vspace{-0.15cm}
\begin{algorithm}	
	\caption{Overall System Algorithm}
	\label{algo_SystemOverallAlgo}
	\fontsize{8pt}{8pt}\selectfont	
	\KwData{$\{L,Q\}$ where $i^{t}=1,2,...,N$}
	\Begin{
		preprocess($\{L,Q\}$)\\
		stayPoints = ValidationStayPointAlgorithm($P$)\\
		clusteredData = DBSCAN(stayPoints)\\
		$T_{x}$ = Trajectory(stayPoints, $P$, clusteredData)\\
		timeSync($\{L,Q\}$)\\
		SFEC($\{L,Q\}$) \\
	}	
	\vspace{-0.15cm}
\end{algorithm}
\vspace{-0.25cm}

We first define set of location data collected from the mobile devices as $L=\{P_{1},P_{2},...,P_{N}\}$, where $P$ can be denoted as $P_{i}=\{\varphi_{i} ,\lambda_{i},a_{i},t_{i}\} $.
The latitude and longitude data are represented by $\varphi$ and $\lambda$ respectively, which is part of GPS Information $P_{i}$.
The $t$ represents the time-stamp of data obtained and $a$ denotes the accuracy of GPS module. 

We denote $Q=\{S_1,S_2,...,S_N\}$ as segment of stay points and define each sensor value for each $Q$ as following
$\text{S}_i = \{v_i, a_i,\mathcal{N}_i,\beta_i, \phi_i\} $. 
Each data are partitioned into segments based on the stay points.
The $v_i$ represents the velocity of the user and this is captured using the mobile device's API, and $a_i$ is the accuracy of location data. 
$\mathcal{N}_i$ is the normalized noise intensity, and $\phi_i$ represents the maximum occurred activity within a time segment.

We noticed different mobile devices possess different noise levels due to different hardware component and manufacturer.
In order to generate a fair comparison of noise level across other devices, we will normalized the noise data using the Equation (\ref{eq_noiseNormalize}) below:
\vspace{-0.10cm}
\begin{equation}
\label{eq_noiseNormalize}
\fontsize{9pt}{9pt}\selectfont
\mathcal{N} = 10\left( {{{s - {s_{\min }}} \over {{s_{\max }} - {s_{\min }}}}} \right)
\end{equation}
where $s$ represents the noise amplitude level. The $s_{min}$ and $s_{max}$ represent minimum and maximum noise level captured respectively based on user data at a particular time frame (i.e. one month data).

\vspace{-0.20cm}

\section{Point of Interest Detection}
\label{sec_POIExtraction}
In this section, we discuss the POI extraction module such as stay points detection and data clustering, that group similar POI into cluster.
\vspace{-0.10cm}

\subsection{Point of Interest Clustering}
\label{subsec_POICluster}

$K$-means clustering \cite{hartigan1979algorithm} is one of the common clustering techniques in handling numerical data since it offers fast and easy usage compared to others.
However, finding an optimal $k$ is one of the challenges as it is an NP hard problem. 
In \cite{laptev2005space}, they have studied the use of $k$-means clustering on spatial and temporal data for POI clustering. 
However, $k$-means is not capable of identifying arbitrary shape of cluster, which makes it not ideal for our proposed architecture.
Alternate common clustering method is density based clustering such as DBSCAN \cite{ester1996density} and OPTICS \cite{ankerst1999optics}. 
In \cite{kisilevich2010p}, they have used customized DBSCAN algorithm for geo-tagging photo which uses density threshold and adaptive density. However, their proposed method only concentrates on spatial information and ignores the temporal data.

Low ultra sampling data collection has a high impact on the quality of data, which may results inaccurate data clustering and wrong POI generated. 
Therefore, we studied different kind of clustering technique as well as stay point detection to ensure best features extraction technique is chosen.
\vspace{-0.10cm}

\subsection{Validation based Stay Points Detection (VSPD) Algorithm}
\label{subsec_clustering}
VSPD algorithms utilize the stay point detection algorithm first proposed in \cite{li2008mining} and we added validation function to remove incompetent POI.
The validation function in VSPD algorithm has been deployed in order to adapt ultra low sampling architecture, which is presented in Algorithm \ref{algo_StayPointClustering}.
\vspace{-0.25cm}
\begin{algorithm}	
	\caption{Validation based Stay Point (VSPD) algorithm}
	\label{algo_StayPointClustering}
	\fontsize{8pt}{8pt}\selectfont
	
	\KwData{$P_i$ where $i=1,2,...,N$}
	\KwResult{$SP_x$}
	\BlankLine
	\For{$k \leftarrow 1 $ to $N$}{
		\While{$j \leftarrow$ ($k+1$)  to $N$}{
			Check($a_{\{i,i+1,...,j\}}$, threshold($a$)) using Equation (\ref{eq_epsCalculate})\\
			Calculate $d$ using Equation (\ref{eq_distance}) \\
			\If{$d > a$}{
				Caclulate $\Delta t$ using Equation (\ref{eq_timeDifference}) \\
				\eIf{$\Delta t > $ threshold($t$)}{
					\If{Validity($d, \Delta t$)}{
						addStayPoint($P_i,P_j$)
					}
				}{break}	
			}
		}
	}
	\vspace{-0.15cm}
\end{algorithm}
\vspace{-0.25cm}

Before going into detail of the algorithm, we first define the following key information, which are stay points and trajectory.
Stay points are defined as $SP$, where it represents potential POI a particular user is at. 
Each stay point can be arranged to trajectory by adding time-stamp, resulting $T_{x} = \{SP_{1},SP_{2},...,SP_{x}\}$.

Reachability of distance between two points is a crucial piece of information required for DBSCAN in order to correctly cluster similar POI. 
Throughout this paper, the distance between two points $\{\varphi_{1},\lambda_{1}\}$, and $\{\varphi_{2},\lambda_{2}\}$ are calculated using the Harvesine formula as shown in Eqn. (\ref{eq_haversineEq}):
\vspace{-0.20cm}
\begin{equation}
\label{eq_haversineEq}
\fontsize{9pt}{9pt}\selectfont
\text{hav}(\dfrac{d}{r})=\text{hav}(\varphi_{2}-\varphi_{1})+\cos(\varphi_{1})\cos(\varphi_{2})\text{hav}(\lambda_{2}-\lambda_{1})
\end{equation}
where harvesine function $hav\text{()}$ can be found in \cite{veness2010calculate} and $d$ is the distance between two points and  $r$ is the radius of earth. The $hav\text{()}$ used in Eqn.(\ref{eq_havOriEq}) is defined as follows:
\begin{equation}
\label{eq_havOriEq}
\fontsize{9pt}{9pt}\selectfont
\text{hav}(\theta) = \sin^{2}(\dfrac{\theta}{2})=\dfrac{1-cos(\theta)}{2}
\end{equation} 
To obtain the distance between two points, we apply the inverse Harvesine function for Equation (\ref{eq_distance}) by using arcsine function as follows:\vspace{-0.03in}
\begin{equation}
\label{eq_distance}
\fontsize{9pt}{9pt}\selectfont
d = r\times\text{hav}^{-1}(h)
= 2r\arcsin \left( {\sqrt h } \right)
\end{equation}\vspace{-0.03in}
After obtaining the distance between two points, we can formulate duration difference between two stay points $\Delta t$ as follows:
\vspace{-0.03in}
\begin{equation}
\label{eq_timeDifference}
\fontsize{9pt}{9pt}\selectfont
\Delta t = t_{i+1} - t_{i}
\end{equation}
where $t_{i}$ is the time-stamp for current sample and $t_{i+1}$ is the timestamp for the previous sample.

Validation function is as a core function in detecting valid stay points.
Ultra low sampling may generate high noise data from environment such as indoor buildings, tunnel, and underground. 
To eliminate potential noise accounted into stay points, we formulate the validation function as follows:

\vspace{-0.03in}
\begin{equation}
\label{eq_validateFunciton}
\fontsize{9pt}{9pt}\selectfont
\text{validaty}(d, \Delta t) = \Delta t < \Theta t \text{ and } d < \Theta d
\end{equation}
\vspace{-0.03in}
where $\Theta t$ is the threshold for time and $\Theta d$ is the threshold for distance.

Subsequently, the $eps$ and $minPts$ need to be determined before applying DBSCAN algorithm. 
The $eps$ are derived using the accuracy of GPS location as follows:
\vspace{-0.04in}
\begin{equation}
\label{eq_epsCalculate}
\fontsize{9pt}{9pt}\selectfont
eps=\left\{\begin{matrix}
a_{i}+a_{i-1}    & \text{if  } a_{i}+a_{i-1}<\Theta l   \text{ where } minPts = 1\\ 
\Theta l & \text{if  } a_{i}+a_{i-1}\ge \Theta l  \text{ where } minPts = 1
\end{matrix}\right.
\end{equation}
where accuracy of GPS data are calculated is considered for checking the validity of the stay points.
For example, if an user entered a tunnel and GPS points remain fixed until he/she exit the tunnel after t time. Our validation model in Eqn. \ref{eq_validateFunciton} will deem the period of entering tunnel as part of traveling because the time taken and distance traveled is considered not valid. 
In Eqn.(\ref{eq_epsCalculate}), a threshold of $\Theta l = 200$ was included to prevent GPS accuracy become too large and form a potential cluster. 
The $minPts$ is set to 1 since we do not treat any POI as outliers at the moment.

\subsection{Evaluation of Stay Point Detection and Clustering}
\label{subsec_locationClusterExp}
To evaluate the proposed VSPD algorithm, we have few volunteers installed the application on their smart phone and proceed with their daily life.
After data collection performed, we will request ground truth from every participant in order to correctly identify the POI they have been to. 
\vspace{-0.15cm}

\subsection*{\textbf{\underline{Effect of Different Clustering Algorithms}}}
\vspace{-0.15cm}
\begin{figure}[ht]
	\centering
	\includegraphics[height=1.2in,width=3.28in]{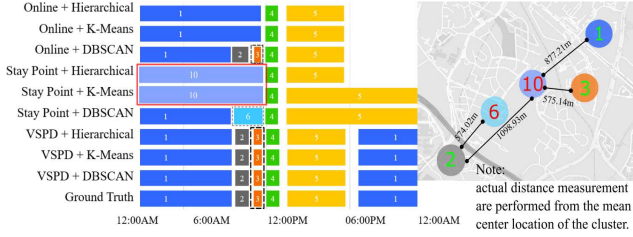}
	\vspace{-1.0em}
	\centering \caption{Comparison of Different Stay Points Extraction and Clustering Techniques}
	\label{fig_diffExtractionClustering}
	\vspace{-0.35cm}
\end{figure} 

First, we study the POI detection algorithm, which consists of (1) VSPD algorithm, (2) stay point algorithm \cite{li2008mining} and (3) online clustering \cite{viswanath2014smart}.
Next, We evaluate the clustering method for each stay point detection algorithm using (1) DBSCAN, (2) K-Means, (3) Hierarchical.
It should be noted, that the $k$-means clustering's $k$ is determined using Davies Bouldin index.
The comparison of different stay point extraction algorithm is presented in Figure \ref{fig_diffExtractionClustering}.

Based on the results, VSPD algorithm and DBSCAN managed to extract POI correctly compared to others. 
Original stay point algorithm clustering wrong results are shown in red solid lined box. 
It is unable to detect the POI correctly as it is unable to distinguish the stay points due to poor indoor GPS accuracy. 
We manually inspect the GPS points located in the cluster 10 and found out that indoor low accuracy led to incorrect stay points extraction.
It is observed that cluster 6 provide incorrect POI compared to ground truth and we highlighted using dotted grey box.
By comparing the different clustering methods, stay point extraction is more crucial in term of extracting the correct POI from the raw database. 
The clustering technique used after stay point extraction algorithm yields significant changes afterwards.
However, by comparing the consistency of the algorithm, we prefer DBSCAN over other clustering techniques due to its capabilities of forming arbitrary shape cluster, as the data collected is never in a consistent shape to begin with.
\vspace{-0.15cm}

\subsection*{\textbf{\underline{Effect of Different Clustering Parameters}}}
\vspace{-0.10cm}

\begin{figure}[h]
	\centering
	\includegraphics[height=1.2in,width=3.48in]{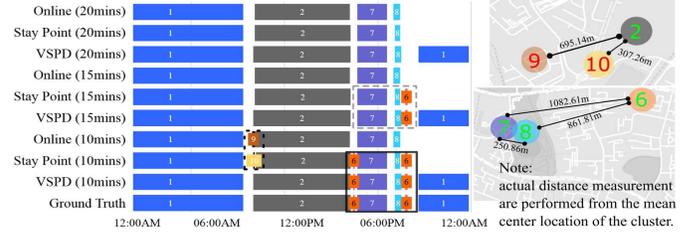}
	\vspace{-1.0em}
	\centering \caption{Comparison of different time threshold for stay point extraction algorithm}
	\label{fig_ClusteringTime}
	\vspace{-0.15cm}
\end{figure}

The next data extraction depends on the duration of stay time at a particular POI. We adjust the delta t accordingly to examine the granulated detail of possible stay point. 
The highlighted area of solid black box and grey dash lined box in Figure \ref{fig_ClusteringTime} generates correct POI according to the ground truth.

However, there is a slight error as framed in black dotted box which is due to transition of GPS accuracy from high to low. 
This causes a gap between two point and the algorithm treat it as two different location and hence error occurred.
After verifying with the volunteers, the reason for such occurrence is die to traffic congestion. 
This led the algorithm to think the users is at a fixed position as it the time does not exceed the $\Delta_t$. 
They appear to be invalid in the validation function in VSPD algorithm and does not form a cluster.

\vspace{-0.15cm}
\vspace{-0.35cm}

\subsection*{\textbf{\underline{Effect of Different Mobile Devices}}}
\begin{figure}[hb]
	\centering
	\includegraphics[height=1.03in,width=3.28in]{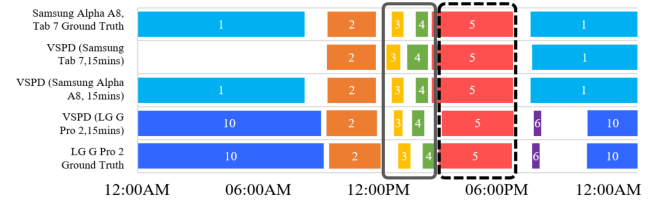}
	\vspace{-1.0em}
	\centering \caption{Data collected using different Mobile Devices}
	\label{fig_DifferentUser}
	\vspace{-0.15cm}
\end{figure}

\vspace{-0.10cm}
Data collection using different smart phone such as Samsung Tab 7, Galaxy Alpha A8 	, LG G-Pro 2 are tested. 
Do note that this is the first phase of our works, we have deployed the application and collected the data more than 100 users in our next works.
In \ref{fig_DifferentUser} same POI are labeled and colored as same number, where other locations are denoted differently.
Slight drift of POI is observed through all devices tested in the same POI as highlighted in Grey solid box and black dash lined box.
Despite slight difference in check-in and check-out time, VSPD algorithm and DBSCAN managed to extract the POI correctly regardless of device vendor.
\vspace{-0.15cm}
The VSPD algorithm has the computational complexity of $O(N^2)$ where validation function is O(1) step with no extra computational cost and is able to eliminate wrong stay points.
\vspace{-0.15cm}

\section{Sensor Fusion Based Environment Classification}
\label{sec_SFEC}

\vspace{-0.15cm}

\subsection{Indoor and Outdoor Classification}
\label{subsec_inOutClassification}
\vspace{-0.10cm}

IO classification have been studied by applying several different approaches. It is can be divided into two approaches, which uses image processing and sensor fusion.

In \cite{li2015iodetector}, IODetector has been proposed to classify IO environment using three sensors (cellular module, light sensor, and magnetic sensor). 
They have combined sensors and wireless approximation methods to determine whether the environment belongs to indoor, outdoor out corridor. 
They have utilized HMM in conjunction with Viterbi algorithm to estimate the IO condition. 
However, Radu et al. \cite{radu2014semi} claimed that their supervised methods are far superior by introducing more features into IO classification. 
They have applied co-training methods and managed to achieve 92\% of accuracy compared to GPS and IODetector. 
Other related work can be found in \cite{bisio2015simple, zou2016bluedetect}.

In \cite{zhang2002boosting}, they have proposed IO classification using image processing. 
Waleed T. et al \cite{tahir2015indoor} has presented classification technique that uses GIST descriptor in conjunction with neural network classifiers to decide whether image is taken indoor or outdoor. Based on their proposed method, they managed to achieve 90.8\% of accuracy by extracting GIST features and using feed-forward neural network to feed training samples.

Since image processing requires training and a camera to be turned on during data collection, it will end up with high battery consumption. Hence, it is not suitable for ultra low sampling rate. Considering the case, mobile device is in pocket, it might end up in inaccurate representation of the IO. To overcome such scenario, we applied the notion of sensor fusion for data collection, in order to achieve better battery management and more accurate environment classification.
\vspace{-0.15cm}

\subsection{Sensor Fusion based Environment Classification (SFEC)}
\label{subsec_IOClassification}
\vspace{-0.10cm}
The stay points are classified as indoor vs. outdoor and private vs. public, using sensor fusion based environment classification. An indoor environment is where the average GPS accuracy for the POI is above a certain threshold, while an outdoor environment is where the average GPS accuracy for the POI is below that threshold (here, the GPS accuracy is returned by Android API, in which, when a higher value for accuracy is returned, it means the GPS accuracy is lower). A private environment is where the noise level for the POI is below a certain threshold and, while a public environment is where the noise level for the POI is above that threshold. Since we cannot obtain a $100\%$ estimation to determine the type of environment where the POI is, we assign percentage confidence levels for determination.

The classifier requires multi-sensor data (i.e. GPS accuracy, Noise, Battery Level, Light etc.), from start and end time of the POI. The duration of POI is divided into 5 minute slots, and each of those slots are given a confidence percentage for type of above mentioned environments. Each type of environment is labeled into one of the 4 different categories, such as indoor, outdoor, private, and public, which are encoded into $\{1,2,3,4\}$ respectively.

Total percentage confidence level for a particular type of environment, for each type of slot, is calculated using the Equation (\ref{eq_sumPercentage}), where, $n$ is number of $5$ minute slots in POI, $P_c$ is confidence percentage, $S^c_k$ is percentage of $k^\text{th}$ slot being type $c$, and $c$ is type of environment where $c=\{1,2,3,4\}$, and $(1\leq k \leq n)$. If there is no data in a slot for the classification, we use $P_0$ to indicate the environment type 'Unclassified'.
\vspace{-0.10cm}
\begin{equation}
\label{eq_sumPercentage}
\fontsize{9pt}{9pt}\selectfont
{P_c} = {1 \over n} \times \sum\limits_{k = 1}^n {S^c_k} \text{\qquad ; if $n > 0$}  
\vspace{-0.1cm}
\end{equation}
The confidence level calculation for the environment type (1) Indoor and (2) Outdoor, are presented in the Eqn. (\ref{eq_IndoorCalculation}) and Eqn. (\ref{eq_OutdoorCalculation}) respectively. 

For $P_1$, percentage contributions from sensors are, $90\%$ by GPS accuracy($G$), $5\%$ by battery level ($\beta$), where $\beta=1$ if battery is charging, and $\beta=0$ otherwise, and $5\%$ by Activity = 'Still' (denoted by $\alpha_s$, where $\alpha_s=\{0,1\}$) which is returned by location API. In Eqn.(\ref{eq_IndoorCalculation}) and Eqn.(\ref{eq_OutdoorCalculation}), $Th_G$ is threshold GPS accuracy, and $x$ is average GPS accuracy in the slot.

\vspace{-0.20cm}
\begin{equation}
\label{eq_IndoorCalculation}
\fontsize{9pt}{9pt}\selectfont
{P_1}=\Bigg[\Big({x-Th_G\over Th_G}\times{0.9}\Big)+ \Big((\beta +\alpha_s)\times 0.05\Big)\Bigg]\text{; if $x>Th_G$}
\vspace{-0.05cm}
\end{equation}

For $P_2$, percentage contributions from sensors are, $90\%$ by $G$, and $10\%$ by light level ($l$), where $l=1$ if light level is above threshold level $Th_l$ or $l=0$ otherwise. $Th_G=30$, $Th_N=5$, and $Th_l=1000$ based on empirical studies. 
\vspace{-0.20cm}
\begin{equation}
\label{eq_OutdoorCalculation}
\fontsize{9pt}{9pt}\selectfont
{P_2}=\Bigg[\Big({Th_G-x\over Th_G}\times{0.9}\Big)+ \Big(Th_l\times 0.1\Big)\Bigg]\text{; if $x<Th_G$}
\end{equation}

The confidence level calculation for the environment type (2) Private and (3) Public, are presented in Eqn. (\ref{eq_PrivateCalculation}) and (\ref{eq_PublicCalculation}) respectively where, $Th_N$ is threshold noise level, and $y$ is average normalized noise level in the slot.

For $P_3$, percentage contributions from sensors are, 90\% by noise level, and 10\% by Activity = \lq Still\rq ($\alpha_s$).
\vspace{-0.20cm}
\begin{equation}
\label{eq_PrivateCalculation}
\fontsize{9pt}{9pt}\selectfont
{P_3}=\Bigg[\Big({Th_N-y\over Th_N}\times{0.9}\Big)+ \Big(\alpha_s\times 0.1\Big)\Bigg]\text{; if $y<Th_N$}
\end{equation}

For $P_4$, percentage contributions from sensors are, 90\%
by noise level, and 10\% by Activity = \lq Walking\rq  (denoted by αw, where $\alpha_w={0,1}$).
\vspace{-0.20cm}
\begin{equation}
\label{eq_PublicCalculation}
\fontsize{9pt}{9pt}\selectfont
{P_4}=\Bigg[\Big({y-Th_N\over Th_N}\times{0.9}\Big)+ \Big(\alpha_w\times 0.1\Big)\Bigg]\text{; if $y>Th_N$}
\end{equation} 

\subsection{Evaluation of SFEC}
\label{subsec_locationClassifyExp}

SFEC is evaluated in real-world scenario by collecting multi-sensor information from mobile device of a volunteer user. Table \ref{table_RGIOResult} shows calculated percentage confidence levels of 4 POI along with their ground truth. When there is no data for a particular environment type, it is labeled as \lq -\rq. For POI $1$, $P_2$ and $P_4$ have substantial difference when compared to $P_1$ and $P_3$ respectively. Similarly, for POI $2$, $P_1$ and $P_3$ have substantial difference in comparison with $P2$ and $P_4$ respectively. For POI $3$, $P_1$ and $P_4$ have substantial difference in contrast with $P_2$ and $P_3$ respectively. For POI $4$, no noise data found possibly due to power management profile of device which causes absence of data and indoor outdoor classification turned out to be incorrect. Moreover, $P_1$ and $P_2$ have no substantial difference to distinguish the environment type. The environment of all POI (except POI $4$) were able to estimate correctly, using SFEC method.   
However, by comparing the P1 and P2 values of POI4, they are closed to each another, this may give us hint that the estimation could be in error.

\begin{table}[t]
	\centering
	\caption{Confidence percentages for SFEC}
	\label{table_RGIOResult}
	\begin{tabular}{@{}lcccc@{}}
		\toprule
		\textbf{Classification}                                                                & \textbf{POI 1}                                                                & \textbf{POI 2}                                                                 & \textbf{POI 3}                                                               & \textbf{POI 4}                                                                \\ \midrule
		P1 (Indoor)                                                                            & -                                                                             & \cellcolor[HTML]{67FD9A}83.94                                                  & \cellcolor[HTML]{67FD9A}59.78                                                & \cellcolor[HTML]{FD6864}35.28                                                 \\
		P2 (Outdoor)                                                                           & \cellcolor[HTML]{67FD9A}60                                                    & 2.42                                                                           & 16.24                                                                        & 23.94                                                                         \\
		P3 (Private)                                                                           & -                                                                             & \cellcolor[HTML]{67FD9A}22.35                                                  & 0.61                                                                         & -                                                                             \\
		P4 (Public)                                                                            & \cellcolor[HTML]{67FD9A}73.12                                                 & -                                                                              & \cellcolor[HTML]{67FD9A}29.66                                                & -                                                                             \\ \midrule
		\multicolumn{1}{|l|}{\textbf{Estimation}}                                              & \multicolumn{1}{c|}{\begin{tabular}[c]{@{}c@{}}Outdoor\\ Public\end{tabular}} & \multicolumn{1}{c|}{\begin{tabular}[c]{@{}c@{}}Indoor\\ Private\end{tabular}}  & \multicolumn{1}{c|}{\begin{tabular}[c]{@{}c@{}}Indoor\\ Public\end{tabular}} & \multicolumn{1}{c|}{\begin{tabular}[c]{@{}c@{}}Indoor\\ Unknown\end{tabular}} \\ \midrule
		\multicolumn{1}{|l|}{\textbf{\begin{tabular}[c]{@{}l@{}}Ground \\ Truth\end{tabular}}} & \multicolumn{1}{c|}{\begin{tabular}[c]{@{}c@{}}Outdoor\\ Public\end{tabular}} & \multicolumn{1}{c|}{\begin{tabular}[c]{@{}c@{}}Indoor \\ Private\end{tabular}} & \multicolumn{1}{c|}{\begin{tabular}[c]{@{}c@{}}Indoor\\ Public\end{tabular}} & \multicolumn{1}{c|}{\begin{tabular}[c]{@{}c@{}}Outdoor\\ Public\end{tabular}} \\ \bottomrule
	\end{tabular}
	\vspace*{-\baselineskip}
\end{table}
\vspace{-0.15cm}

\section{Conclusion}
\label{sec_conclusion}
\vspace{-0.138cm}

In a nutshell, this paper focuses on low sampling and sensor fusion techniques for POI extraction and environment classification for smart-phone location data.
We introduced validation function for the POI extraction and clustered identical POI to generate user trajectory.
Furthermore, we implemented SFEC method to identify the type of environment using various sensor input.
In future, we will expand the analysis model to study regional mobility and travel pattern with a large scale crow sourcing / sensing activity, where ground truth is hard to obtain. 

\vspace{-0.20cm}

\section*{Acknowledgment}
Authors would like to express their sincere gratitude to all the participants in the study who allowed us to collect their location information. This research is supported by the Lee Kuan Yew Centre for Innovative Cities under Lee Li Ming Programme in Aging Urbanism.
\vspace{-0.20cm}

\bibliographystyle{IEEEtran}
\newcommand{\BIBdecl}{\setlength{\itemsep}{0.22 em}}
\bibliography{BibSpace}  

\begin{thebibliography}{10}
\providecommand{\url}[1]{#1}
\csname url@samestyle\endcsname
\providecommand{\newblock}{\relax}
\providecommand{\bibinfo}[2]{#2}
\providecommand{\BIBentrySTDinterwordspacing}{\spaceskip=0pt\relax}
\providecommand{\BIBentryALTinterwordstretchfactor}{4}
\providecommand{\BIBentryALTinterwordspacing}{\spaceskip=\fontdimen2\font plus
\BIBentryALTinterwordstretchfactor\fontdimen3\font minus
  \fontdimen4\font\relax}
\providecommand{\BIBforeignlanguage}[2]{{%
\expandafter\ifx\csname l@#1\endcsname\relax
\typeout{** WARNING: IEEEtran.bst: No hyphenation pattern has been}%
\typeout{** loaded for the language `#1'. Using the pattern for}%
\typeout{** the default language instead.}%
\else
\language=\csname l@#1\endcsname
\fi
#2}}
\providecommand{\BIBdecl}{\relax}
\BIBdecl

\bibitem{hoteit2014estimating}
S.~Hoteit, S.~Secci, S.~Sobolevsky, C.~Ratti, and G.~Pujolle, ``Estimating
  human trajectories and hotspots through mobile phone data,'' \emph{Computer
  Networks}, vol.~64, pp. 296--307, 2014.

\bibitem{kang2013exploring}
C.~Kang, S.~Sobolevsky, Y.~Liu, and C.~Ratti, ``Exploring human movements in
  singapore: a comparative analysis based on mobile phone and taxicab usages,''
  in \emph{Proceedings of the 2nd ACM SIGKDD international workshop on urban
  computing}.\hskip 1em plus 0.5em minus 0.4em\relax ACM, 2013, p.~1.

\bibitem{viswanath2014smart}
S.~K. Viswanath, C.~Yuen, X.~Ku, and X.~Liu, ``Smart tourist-passive mobility
  tracking through mobile application,'' in \emph{International Internet of
  Things Summit}.\hskip 1em plus 0.5em minus 0.4em\relax Springer, 2014, pp.
  183--191.

\bibitem{li2008mining}
Q.~Li, Y.~Zheng, X.~Xie, Y.~Chen, W.~Liu, and W.-Y. Ma, ``Mining user
  similarity based on location history,'' in \emph{Proceedings of the 16th ACM
  SIGSPATIAL international conference on Advances in geographic information
  systems}.\hskip 1em plus 0.5em minus 0.4em\relax ACM, 2008, p.~34.

\bibitem{lou2009map}
Y.~Lou, C.~Zhang, Y.~Zheng, X.~Xie, W.~Wang, and Y.~Huang, ``Map-matching for
  low-sampling-rate gps trajectories,'' in \emph{Proceedings of the 17th ACM
  SIGSPATIAL International Conference on Advances in Geographic Information
  Systems}.\hskip 1em plus 0.5em minus 0.4em\relax ACM, 2009, pp. 352--361.

\bibitem{zheng2008learning}
Y.~Zheng, L.~Liu, L.~Wang, and X.~Xie, ``Learning transportation mode from raw
  gps data for geographic applications on the web,'' in \emph{Proceedings of
  the 17th international conference on World Wide Web}.\hskip 1em plus 0.5em
  minus 0.4em\relax ACM, 2008, pp. 247--256.

\bibitem{zheng2010geolife}
Y.~Zheng, X.~Xie, and W.-Y. Ma, ``Geolife: A collaborative social networking
  service among user, location and trajectory.'' \emph{IEEE Data Eng. Bull.},
  vol.~33, no.~2, pp. 32--39, 2010.

\bibitem{suzuki2007learning}
N.~Suzuki, K.~Hirasawa, K.~Tanaka, Y.~Kobayashi, Y.~Sato, and Y.~Fujino,
  ``Learning motion patterns and anomaly detection by human trajectory
  analysis,'' in \emph{2007 IEEE International Conference on Systems, Man and
  Cybernetics}.\hskip 1em plus 0.5em minus 0.4em\relax IEEE, 2007, pp.
  498--503.

\bibitem{hartigan1979algorithm}
J.~A. Hartigan and M.~A. Wong, ``Algorithm as 136: A k-means clustering
  algorithm,'' \emph{Journal of the Royal Statistical Society. Series C
  (Applied Statistics)}, vol.~28, no.~1, pp. 100--108, 1979.

\bibitem{laptev2005space}
I.~Laptev, ``On space-time interest points,'' \emph{International Journal of
  Computer Vision}, vol.~64, no. 2-3, pp. 107--123, 2005.

\bibitem{ester1996density}
M.~Ester, H.-P. Kriegel, J.~Sander, X.~Xu \emph{et~al.}, ``A density-based
  algorithm for discovering clusters in large spatial databases with noise.''
  in \emph{Kdd}, vol.~96, no.~34, 1996, pp. 226--231.

\bibitem{ankerst1999optics}
M.~Ankerst, M.~M. Breunig, H.-P. Kriegel, and J.~Sander, ``Optics: ordering
  points to identify the clustering structure,'' in \emph{ACM Sigmod Record},
  vol.~28, no.~2.\hskip 1em plus 0.5em minus 0.4em\relax ACM, 1999, pp. 49--60.

\bibitem{kisilevich2010p}
S.~Kisilevich, F.~Mansmann, and D.~Keim, ``P-dbscan: a density based clustering
  algorithm for exploration and analysis of attractive areas using collections
  of geo-tagged photos,'' in \emph{Proceedings of the 1st international
  conference and exhibition on computing for geospatial research \&
  application}.\hskip 1em plus 0.5em minus 0.4em\relax ACM, 2010, p.~38.

\bibitem{veness2010calculate}
C.~Veness, ``Calculate distance, bearing and more between latitude/longitude
  points,'' \emph{not dated, http://www. movable-type. co. uk/scripts/latlong.
  html}, 2010.

\bibitem{li2015iodetector}
M.~Li, P.~Zhou, Y.~Zheng, Z.~Li, and G.~Shen, ``Iodetector: A generic service
  for indoor/outdoor detection,'' \emph{ACM Transactions on Sensor Networks
  (TOSN)}, vol.~11, no.~2, p.~28, 2015.

\bibitem{radu2014semi}
V.~Radu, P.~Katsikouli, R.~Sarkar, and M.~K. Marina, ``A semi-supervised
  learning approach for robust indoor-outdoor detection with smartphones,'' in
  \emph{Proceedings of the 12th ACM Conference on Embedded Network Sensor
  Systems}.\hskip 1em plus 0.5em minus 0.4em\relax ACM, 2014, pp. 280--294.

\bibitem{bisio2015simple}
I.~Bisio, A.~Delfino, and F.~Lavagetto, ``A simple ultrasonic indoor/outdoor
  detector for mobile devices,'' in \emph{2015 International Wireless
  Communications and Mobile Computing Conference (IWCMC)}.\hskip 1em plus 0.5em
  minus 0.4em\relax IEEE, 2015, pp. 137--141.

\bibitem{zou2016bluedetect}
H.~Zou, H.~Jiang, Y.~Luo, J.~Zhu, X.~Lu, and L.~Xie, ``Bluedetect: An
  ibeacon-enabled scheme for accurate and energy-efficient indoor-outdoor
  detection and seamless location-based service,'' \emph{Sensors}, vol.~16,
  no.~2, p. 268, 2016.

\bibitem{zhang2002boosting}
L.~Zhang, M.~Li, and H.-J. Zhang, ``Boosting image orientation detection with
  indoor vs. outdoor classification,'' in \emph{Applications of Computer
  Vision, 2002.(WACV 2002). Proceedings. Sixth IEEE Workshop on}.\hskip 1em
  plus 0.5em minus 0.4em\relax IEEE, 2002, pp. 95--99.

\bibitem{tahir2015indoor}
W.~Tahir, A.~Majeed, and T.~Rehman, ``Indoor/outdoor image classification using
  gist image features and neural network classifiers,'' in \emph{2015 12th
  International Conference on High-capacity Optical Networks and
  Enabling/Emerging Technologies (HONET)}.\hskip 1em plus 0.5em minus
  0.4em\relax IEEE, 2015, pp. 1--5.

\end{thebibliography}

\end{document}